\newcounter{myctr}
\def\myitem{\refstepcounter{myctr}\bibfont\noindent\ifnum\themyctr>9\else\phantom{0}\fi\hangindent17pt\themyctr.\enskip}

\documentclass{ws-ijqi}

\usepackage{amsmath}
\usepackage{graphicx}
\usepackage{epsfig}
\usepackage{color}
\usepackage{dsfont}
\usepackage{pgfplots}
\usepackage{subfigure}

\newcommand{\ket}[1]{\vert #1 \rangle}
\newcommand{\bra}[1]{\langle #1 \vert}

\newcommand{\ketbra}[2]{\vert #1 \rangle \langle #2 \vert}
\newcommand{\imm}{{\rm i }}

\newcommand{\adag}{a^\dag}

\begin{document}

\markboth{M. Bina, S. Maffezzoli Felis, and S. Olivares}
{Entanglement generation in the ultra-strongly coupled Rabi model}

\catchline{}{}{}{}{}

\title{ENTANGLEMENT GENERATION\\ IN THE ULTRA-STRONGLY COUPLED RABI MODEL }

\author{MATTEO BINA}
\address{Dipartimento di Fisica, Universit\`a degli 
Studi di Milano\\ via Celoria 16, I-20133 Milano, Italy\\
matteo.bina@gmail.com}

\author{STEFANO MAFFEZZOLI FELIS}
\address{Dipartimento di Fisica, Universit\`a degli 
Studi di Milano\\ via Celoria 16, I-20133 Milano, Italy}

\author{STEFANO OLIVARES}
\address{Dipartimento di Fisica and CNISM, Universit\`a degli 
Studi di Milano\\ via Celoria 16, I-20133  Milano, Italy\\ stefano.olivares@fisica.unimi.it}

\maketitle


\begin{abstract}
We analyze the dynamics of the quantum Rabi model for two qubits interacting through a common bosonic field mode (resonator), focusing on the generation and detection of maximally entangled Bell states. We obtain analytical results for the unitary dynamics of this system in the slow-qubit (or degenerate) regime, considering ultra-strong coupling between qubits and resonator mode, for which the rotating wave approximation is no longer applicable. We also numerically investigate the dynamics beyond the slow-qubit condition in order to study the validity of the model in the presence of less strict conditions.
\end{abstract}


\section{Introduction}

Light-matter interaction at the fundamental quantum level finds its paradigmatic description in the quantum Rabi model for a two-level system ($\frac12$-spin particle) coupled via dipole-like interaction to a single mode oscillator (bosonic field).\cite{Haroche} An approximated model\cite{JaynesCummings} introduced by Jaynes and Cummings (JC) in 1964 gained great popularity in the cavity quantum electrodynamics (CQED) context due to its wide applicability in countless experiments.\cite{HarocheRabi,Kimble,Wineland}
The rotating wave approximation (RWA) is the cornerstone of the JC model, as it allows to neglect counter-rotating terms in the system Hamiltonian which, otherwise, would produce double excitation (or de-excitation) processes forbidden by the low ratios between the spin-boson couplings.\cite{ShoreKnight,Bina} Outstanding experiments in CQED, ultracold atoms and optomechanical systems always exploited and validated this model,\cite{JCSpecialIssue} until the advent of superconducting circuits and devices which mimic a two-level system strongly interacting with a single bosonic field mode.\cite{Blais,Wallraff,Chiorescu,Clarke,TransmonQubit} In this recent framework, the $\frac12$-spin-like particle (artificial atom) and a microwave field mode of a transmission line resonator may achieve such high interaction strengths to invalidate the RWA, opening new possibilities for the implementation of quantum optics and quantum information processes.\cite{Blais2,Nakamura,Peropadre,Ciuti} Many attempts to obtain approximated analytical solutions for the spectrum of the system have been performed,\cite{prlIrish,prbIrish} but the integrability of the quantum Rabi model has been only recently demonstrated for the whole range of the involved parameters,\cite{prlBraak} also for the case involving two qubits.\cite{JphysMathBraak}
\par
Following this front, in this paper we focus on the unitary dynamics of two qubits ultra-strongly coupled to a single bosonic field mode of a resonator described by the two-qubit quantum Rabi model. In particular we address the so called slow-qubit or degenerate approximation, considering low qubit transition frequencies and, consequently, large detunings. This range of parameters allows us to describe in an analytical way the dynamics of the system and to extract the mechanism underlying the generation of maximally entangled Bell states for the two-qubit subsystem.
\par
The paper is structured as follows. Section~\ref{sec:level1} introduces the Hamiltonian model for large detuning and subsequently we study the unitary dynamics (section~\ref{sec:unitary}), providing an explicit analytical solution in a proper time scale. In section~\ref{sec:bell} we derive the conditions for the generation of maximally entangled states in terms of the system parameters, showing that the reversibility of the time evolution could be used to readout the state of the two qubits. Finally, in section~\ref{sec:numerical} we show numerical results confirming the validity of the approximated model in both cases in which the transition frequencies of the qubits are considered equal and different. Section~\ref{sec:concl} closes the paper drawing some concluding remarks.

\section{\label{sec:level1}Two-qubit Rabi model in large detuning}
\begin{figure}[htb]
\begin{center}
\includegraphics[scale=0.35]{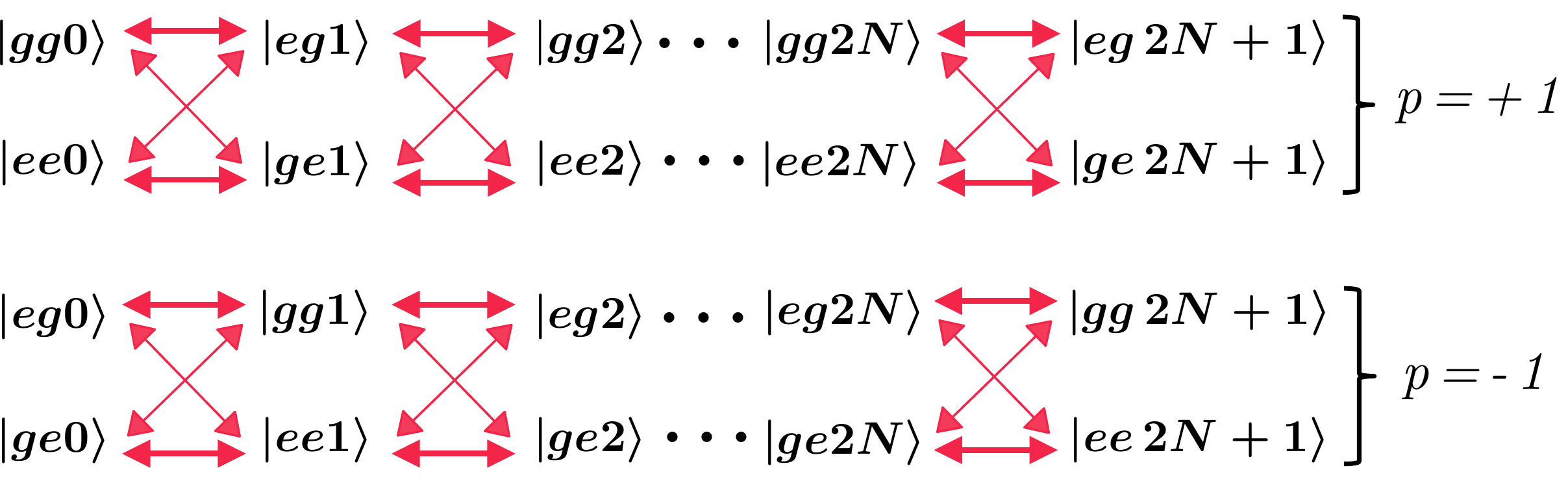}
\caption{(Color online) Schematic view of the two parity chains dividing the whole system Hilbert space. Arrows show the connections among the states of each parity given by the interaction term in Eq. (\ref{HRabi}). See the text for details. \label{fig:ParityChains}}
\end{center}
\end{figure}

We consider two qubits of transition frequencies $\omega_i$ between the states $\ket{e}_i$ and $\ket{g}_i$, $i=1,2$, interacting with a common bosonic field mode, oscillating at the frequency $\omega$
and described by the field operators $a$ and $a^\dag$, $[a,a^\dag]=1$. This system is described by the quantum Rabi model with Hamiltonian given by
\begin{equation}
\label{HRabi}
 H=\hbar\omega\adag a+ \hbar\sum_{i=1,2}\frac{\omega_i}{2}\sigma_z^{(i)}+\hbar \sum_{i=1,2}\Gamma_i \sigma_x^{(i)}\left(\adag+a\right),
\end{equation}
$\Gamma_1$ and $\Gamma_2$ being the coupling constants, $\sigma_z^{(i)}=\ket{e}_i\bra{e}-\ket{g}_i\bra{g}$ and $\sigma_x^{(i)}=\sigma_+^{(i)}+\sigma_-^{(i)}$, where $\sigma_+^{(i)} = \ket{e}_i\bra{g}$ and $\sigma_-^{(i)}=\ket{g}_i\bra{e}$ are the raising and lowering operators, respectively, of the $i$-th qubit. In many contexts, ranging from cavity to circuit QED and from trapped ions to micro-mechanical resonators, the quantum Rabi model can be effectively reduced to the well known JC model applying the RWA, consisting in dropping the counter-rotating high-frequency terms $\sigma_+^{(i)}a$ and $\sigma_-^{(i)}a^\dag$. However, in the so-called ultra-strong ($\Gamma/\omega\gtrsim 10^{-1}$) or deep-strong ($\Gamma/\omega\gtrsim 1$) coupling regime\cite{prlCasanova}, the RWA is no longer applicable an these terms must be included, leading to Eq. (\ref{HRabi}) as the proper Hamiltonian of the considered system. If in the JC model the total number of excitation $N_e=a^\dag a +\sum_i \sigma_+^{(i)}\sigma_-^{(i)}$ was the conserved quantity, in the quantum Rabi model the only constant of motion is the parity $ \Pi_2=\sigma^{(1)}_z\sigma^{(2)}_z(-)^{\adag a}={\rm e}^{\imm \pi N_e}$.
The operator $\Pi_2$ has only two eigenvalues $p=\pm1$. The Hilbert space of the whole system is thus divided in two subspaces spanned by the states $\ket{Q\,n} \equiv \ket{Q}\otimes\ket{n}$, where $\ket{Q} \in \left\{\ket{gg},\ket{ge},\ket{eg},\ket{ee}\right\}$ and $n \in {\mathbb N}$, with even parity ($p=1$ and $N_e$ even) and odd parity ($p=-1$ and $N_e$ odd), giving rise to two unconnected parity chains of states (see Fig. \ref{fig:ParityChains}). The conservation law  $[\Pi_2, H]=0$ associated with the parity operator ensures that the dynamics develops in one of these two Hilbert subspaces.\cite{prlCasanova} 

The integrability and diagonalization of the quantum Rabi model has been largely investigated,\cite{prlBraak,JPABraak} together with analytical approaches to find approximated spectra.\cite{prlIrish} Here, we consider a range of parameters that allows us to analytically describe the dynamics of the system capturing a peculiar and interesting physical behavior of the subsystems. In particular, we describe the dynamics in the so called slow-qubit (or degenerate) regime $\omega_i\ll \omega$, $i=1,2$, for which the transition frequency of the qubits is much smaller than the frequency of the bosonic mode. The large detuning condition is equivalent to describe the system dynamics on a proper time scale, as we explore in the next sections.
\par
From now on, we assume, without loss of generality, equal qubit frequencies $\omega_1=\omega_2\equiv\omega_q$ and introduce the detuning parameter $\Delta\equiv\omega-\omega_q$ with the bosonic field mode. By carrying out a common interaction picture transformation and approximating the oscillating exponential factors at zeroth order for $\omega_q t\ll 1$, we obtain the Hamiltonian
\begin{equation}\label{H':old}
 H'=\hbar \Delta \adag a+ \hbar \sum_{i=1,2}\Gamma_i \sigma_x^{(i)}\left(\adag+a\right),
\end{equation}
which describes a displaced harmonic oscillator dependent on the state of the two qubits. Equation~(\ref{H':old}) can be rewritten in a more compact way as
\begin{equation}\label{H'}
H'=\hbar\Delta\left[D^{\dagger}\left(\gamma\right)\adag a D\left(\gamma\right)-\gamma^2\right],
\end{equation}
where we have introduced the two-qubit operator $$\gamma=\frac{\Gamma_1}{\Delta}\sigma_x^{(1)}+\frac{\Gamma_2}{\Delta}\sigma_x^{(2)},$$ and $D(\gamma)=\exp\{\gamma a^\dag -\gamma^\dag a\}$ is a displacement operator conditioned by the two-qubit state expressed in the rotated basis $\{\ket{++},\ket{+-},\ket{-+},\ket{--}\}$, with $\ket{\pm}_i=(\ket{g}_i\pm\ket{e}_i)/\sqrt{2}$. The operator $\gamma$ is diagonal in the rotated basis and displays four eigenvalues $\{\gamma_+,\gamma_-,-\gamma_+,-\gamma_-\}$, with $\gamma_\pm=(\Gamma_1\pm\Gamma_2)/\Delta$.
It is worth noting that $\gamma^2$ in Eq. (\ref{H'}) contains the term $\sigma_x^{(1)}\sigma_x^{(2)}$, which is at the basis of the entangling mechanism as we will see in the analysis of the subsystem dynamics.

\section{Unitary dynamics}\label{sec:unitary}
We can describe analytically the unitary dynamics of the considered ultra-strongly coupled system only for certain values of parameters satisfying $\omega_i \ll \omega$. 
The time-evolution operator generated by the Hamiltonian (\ref{H'}) can be written as
\begin{align}
U(t)&=
D^{\dagger}(\gamma){\rm e}^{-\imm t\Delta\adag a}D(\gamma)e^{\imm t\Delta\gamma^2},\\
&=e^{\imm \phi_t\gamma^2}D\left(\gamma_t\right)e^{-\imm t\Delta\adag a},\label{U}
\end{align}
with $\gamma_t=({\rm e}^{-\imm t\Delta}-1)\gamma$ and $\phi_t=t\Delta-\sin(t\Delta)$, and where we used $f({\rm e}^{A} B {\rm e}^{-A})={\rm e}^{A} f(B) {\rm e}^{-A}$ derived from the Baker-Campbell-Hausdorff formula.\cite{Barnett} From Eq.~(\ref{U}) we can deduce the dynamics of our system, which is a composition of the following operations: a time- and energy-dependent phase shift, a displacement conditioned by the two-qubit state at time $t$ and the time-dependent entangler operator of the two qubits, resulting into an oscillatory behavior with period $t=2\pi/\Delta$. We point out that in the dynamics described by Eq.~(\ref{U}) the operator $\gamma^2$ brings a global phase $\phi_G(t)$ and a local phase $\phi_L(t)$, namely:
\begin{align}
\phi_G(t)=\phi_t\,\frac{\gamma_+^2 +\,\gamma_-^2}{2}=\phi_t\frac{\Gamma_1^2+\Gamma_2^2}{\Delta^2}, \quad
\phi_L(t)=\phi\,\frac{\gamma_+^2 -\,\gamma_-^2}{2}=\phi_t\frac{2\, \Gamma_1\Gamma_2}{\Delta^2}\label{PhaseL}.
\end{align}
As we will see, while $\phi_G(t)$ is linked to the identity operator, $\phi_L(t)$ is linked to the operator $\sigma_x^{(1)}\sigma_x^{(2)}$, which is responsible for the entangling mechanism.

\section{Entangling two qubits and Bell state analyzer}\label{sec:bell}
If we assume that the initial state of the system is $\ket{\psi(0)}=\ket{gg\,0}$, then the evolved state of the system $\ket{\psi (t)} = U(t)\ket{\psi(0)}$ can be written as
\begin{align}
\ket{\psi &(t)}=
\frac{1}{4} \, \exp\left\{ \imm [\phi_G(t) + \phi_L(t)] \right\} \notag\\
&\times\biggl\{\ket{gg}\Bigl[\ket{c_E^{(+)}(t)}+e^{-2\imm\phi_L(t)}\ket{c_E^{(-)}(t)}\Bigr]
+\ket{ge}\Bigl[\ket{c_O^{(+)}(t)}-e^{-2\imm\phi_L(t)}\ket{c_O^{(-)}(t)}\Bigr]\notag\\
&+\ket{eg}\Bigl[\ket{c_O^{(+)}(t)}+e^{-2\imm\phi_L(t)}\ket{c_O^{(-)}(t)}\Bigr]
+\ket{ee}\Bigl[\ket{c_E^{(+)}(t)}-e^{-2\imm\phi_L(t)}\ket{c_E^{(-)}(t)}\Bigr]
\biggr\},
\label{Psi_t_I}
\end{align}
and we note that for each two-qubit state $\ket{Q}$, there is a peculiar superposition of Schr\"odinger cat-like states  $\ket{c_E^{(\pm)}(t)}\equiv\ket{\gamma_{t,\pm}}+\ket{-\gamma_{t,\pm}}$ (\emph{even} cat state) and $\ket{c_O^{(\pm)}(t)}\equiv\ket{\gamma_{t,\pm}}-\ket{-\gamma_{t,\pm}}$ (\emph{odd} cat state), with $\gamma_{t,\pm}=({\rm e}^{-\imm t\Delta}-1)\gamma_\pm$. Note that though $\ket{c_E^{(\pm)}(t)}$ an $\ket{c_O^{(\pm)}(t)}$ are not normalized, the superpositions appearing in Eq.~(\ref{Psi_t_I}) are normalized states of the resonator.
\par
As our focus is to study the two-qubit entanglement generated by this dynamics, we point out that for $t_n=2\pi n/\Delta$, with $n\in\mathbb{N}$, the resonator mode bounces back into the initial state $\ket{0}$ as $\gamma_\pm(t_n)=0$, and the two-qubit subsystem is left in the superposition
\begin{align}\label{2q_state_tn}
\ket{\psi(t_n)}=
\big \{ \cos\big[\phi_L(t_n)\big]\ket{gg} +
\imm\sin\big[\phi_L(t_n)\big]\ket{ee}\big \}\otimes\ket{0}
\equiv \ket{\psi_q(t_n)}\otimes \ket{0},
\end{align}
Now, by imposing the condition
\begin{equation}\label{CouplingRelation}
\Gamma_1 \Gamma_2=\frac{2m+1}{16n}\, \Delta^2 ,
\end{equation} 
on the coupling parameters $\Gamma_1$ and $\Gamma_2$, with $n,m\in\mathbb{N}$, and inserting it in Eq. (\ref{PhaseL}) with $t_n=2\pi n/\Delta$, we obtain $\phi_L(t_n)=(2m+1)\frac{\pi}{4}$ and Eq. (\ref{2q_state_tn}) reduces to $\ket{\psi^{(+)}}\otimes\ket{0}$, for even $m$, or $\ket{\psi^{(-)}}\otimes\ket{0}$, for odd $m$, where we introduced the Bell states:
\begin{equation}\label{BellStates}
\begin{split}
\ket{\psi^{(\pm)}}&=\frac{1}{\sqrt{2}}\left(\ket{gg}\pm\imm\ket{ee}\right),
\quad
\ket{\phi^{(\pm)}}=\frac{1}{\sqrt{2}}\left(\ket{eg}\pm\imm\ket{ge}\right).
\end{split}
\end{equation}
Therefore, our protocol allows to generate all the four states (\ref{BellStates}) at the first peak time $t_1$ and under the condition (\ref{CouplingRelation}), by choosing the different initial states $\ket{Q\, 0}$, namely:
\begin{equation}\label{BellStatesN}
\begin{split}
U(t_1)\ket{gg\,0}&=\ket{\psi^{(+)}}\otimes\ket{0},
\quad
U(t_1)\ket{ee\,0}=\ket{\psi^{(-)}}\otimes\ket{0},\\
U(t_1)\ket{eg\,0}&=\ket{\phi^{(+)}}\otimes\ket{0},
\quad
U(t_1)\ket{ge\,0}=\ket{\phi^{(-)}}\otimes\ket{0}.
\end{split}
\end{equation}
It is worth noting that in the context of circuit QED,\cite{RomeroGates,Majer,MakhlinReview} or quantum simulation,\cite{BallesterSimulationUSC,BinaPRA} it is possible to suitably tune the parameters to satisfy condition (\ref{CouplingRelation}), thus implementing the entangling protocol.
\par
Conversely, being the unitary dynamics reversible, it is also possible to use this gate as an analyzer of Bell states. Indeed, choosing as the initial state one of the Bell states in Eq. (\ref{BellStates}), the evolved state of the two qubits at the first peak time $t_1$  is one of the factorized bipartite state of the standard basis $\ket{Q\, n}$ which can be measured with standard techniques, obtaining information on the initial Bell state.\\
Furthermore, this gate for the generation of maximally entangled states is robust against the choice of the initial state of the bosonic subsystem. Indeed, it can be easily demonstrated that for every preparation of the resonator mode the oscillatory dynamics is the same as described above: after every time $t=t_n$ the initial state of the resonator is restored and the state of the two qubits is left in the superposition state (\ref{2q_state_tn}). In fact, considering a generic initial state of the bosonic subsystem $\rho_b(0)=\sum_{l,m}\rho_{l,m}\ketbra{l}{m}$ expressed on the Fock basis, and $\ket{gg}$ as initial preparation of the two qubits, then the evolved state at time $t_n$ is still a factorized state of the two subsystems, namely:
\begin{equation}
 \rho(t_n)=\ket{\psi_q(t_n)}\bra{\psi_q(t_n)}\otimes\rho_b(0),
\end{equation}
where $\ket{\psi_q(t_n)}$ is given in Eq.~(\ref{2q_state_tn}).
\begin{figure}[t]
\begin{center}
\includegraphics[scale=0.32]{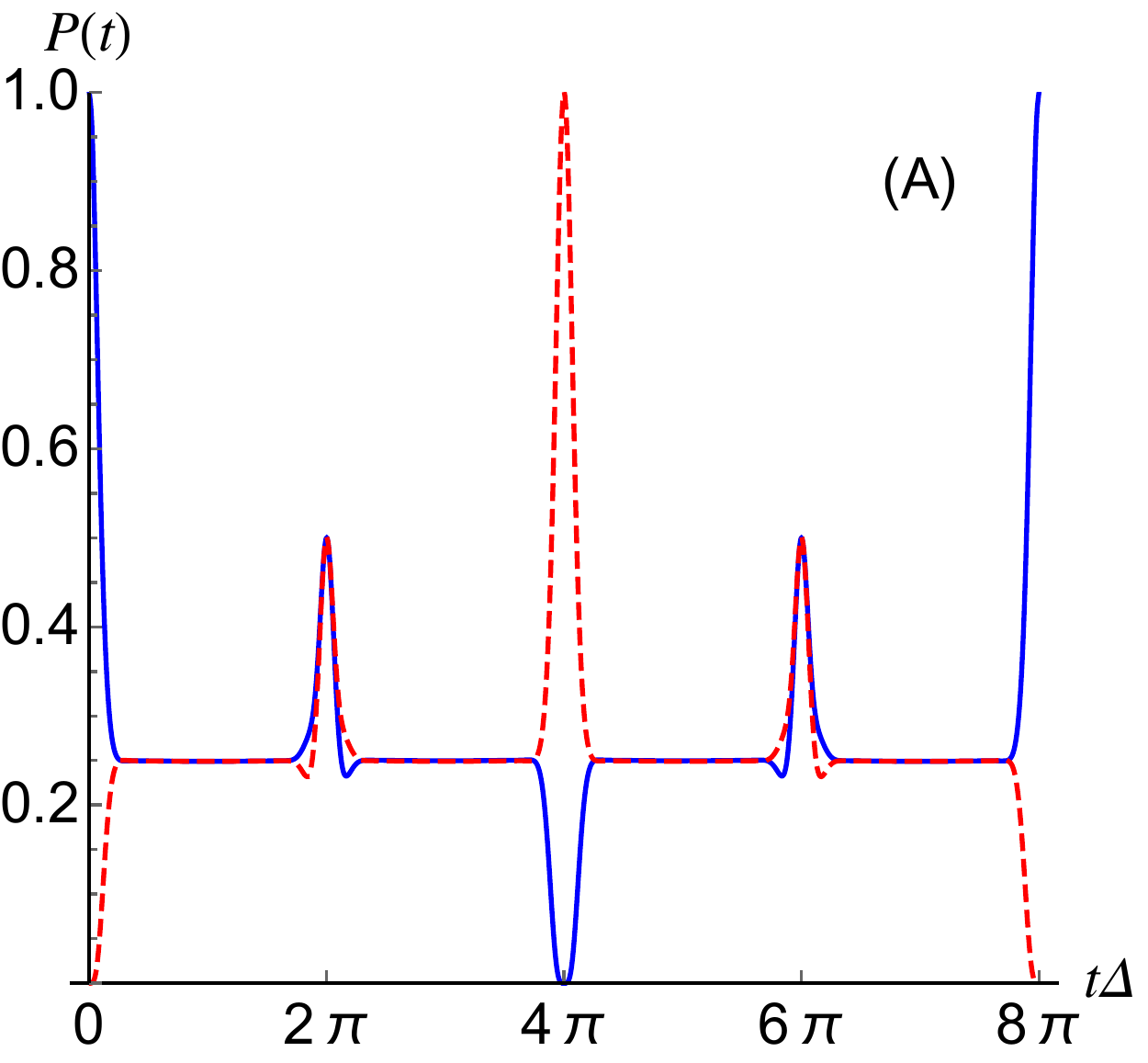}
\includegraphics[scale=0.32]{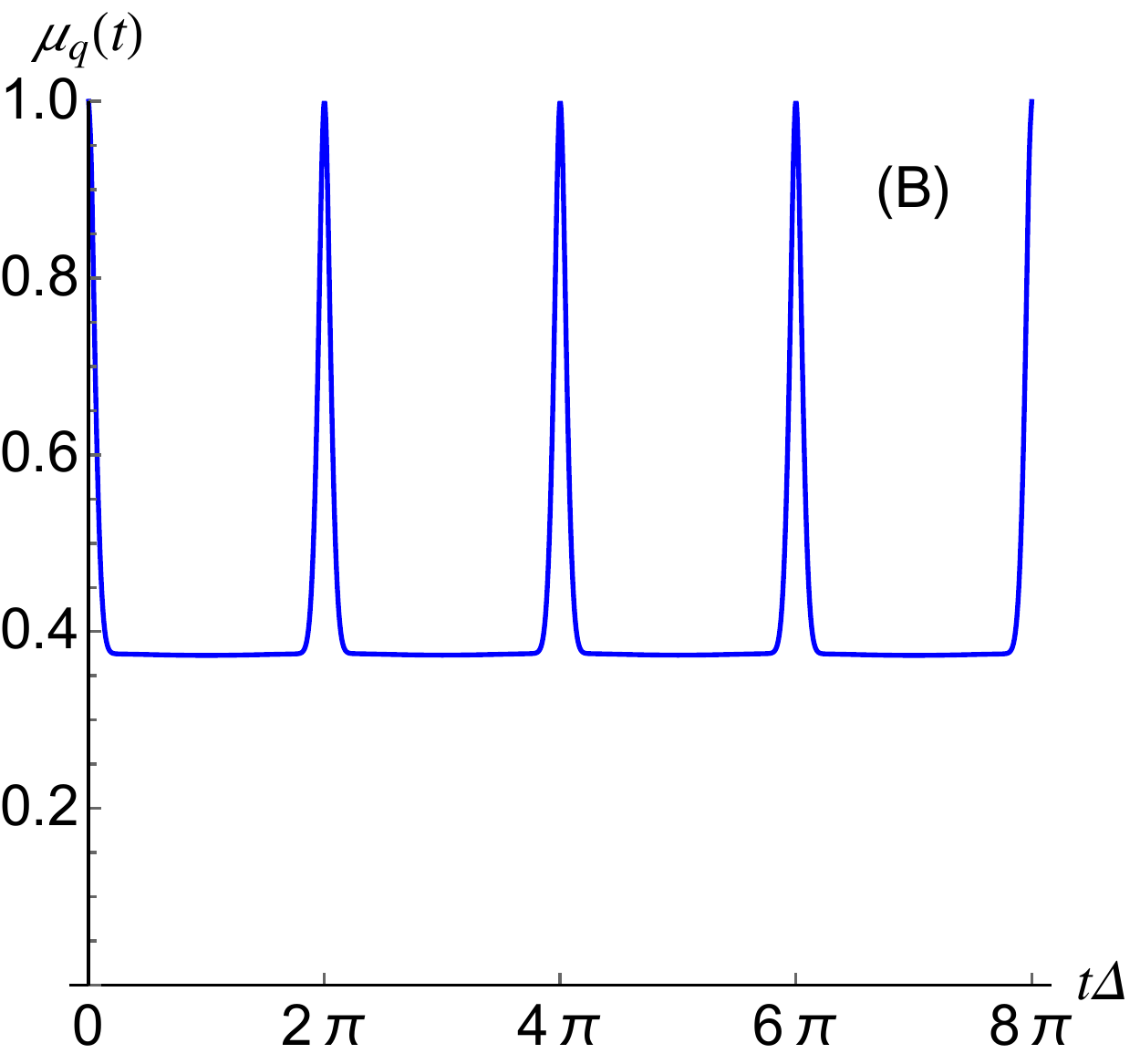}
\includegraphics[scale=0.32]{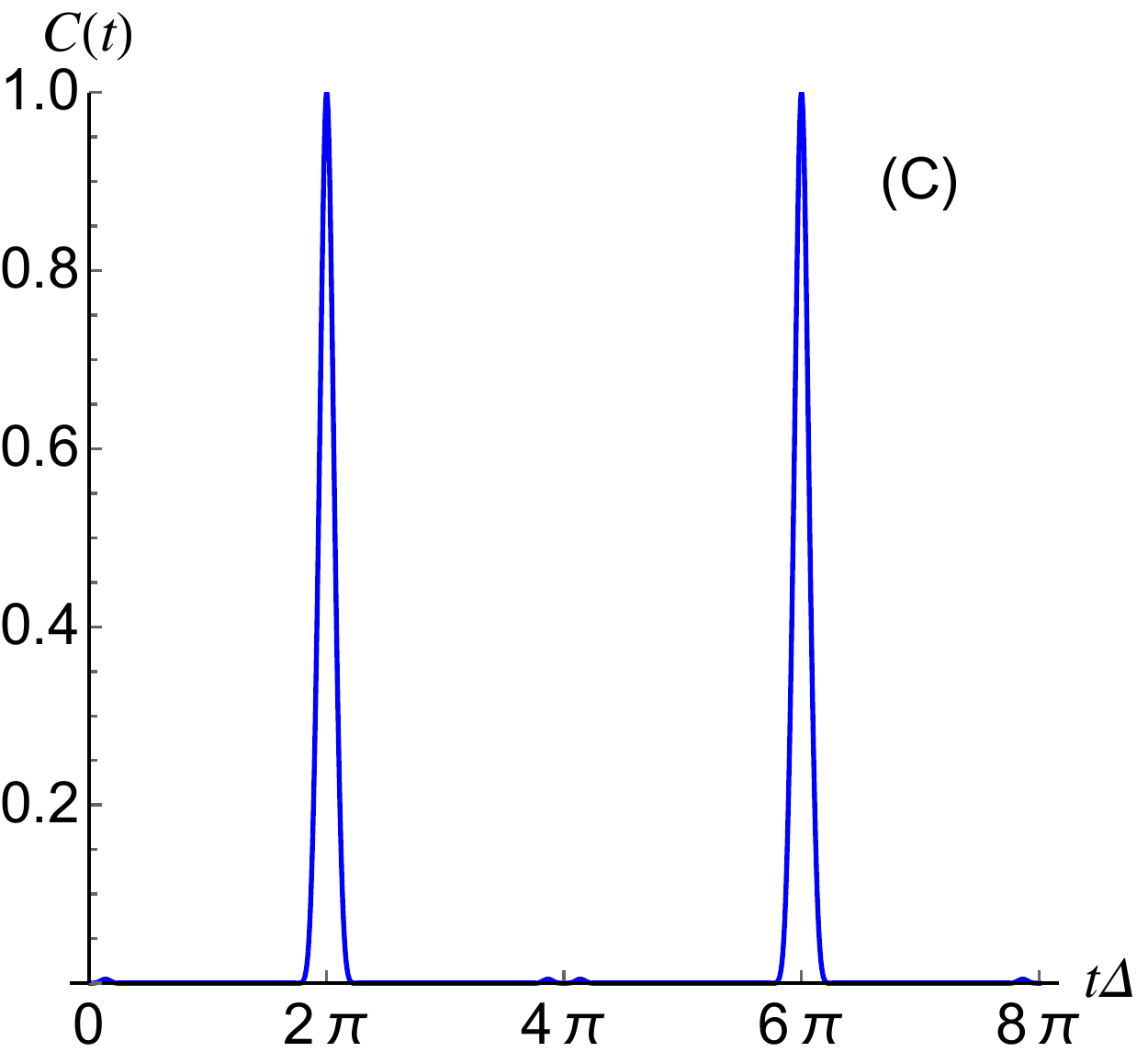}
\caption{(Color online) Dynamics of the system as a function of $t\Delta$ starting from the state $\ket{gg\, 0}$, with parameters  $\Gamma_1=2\Delta$ and $\Gamma_2=\frac{65}{32}\Delta\approx2.03\Delta$ (deep-strong coupling regime) satisfying Eq. (\ref{CouplingRelation}). Panel (A): Joint probability distributions $P_{gg\,0}(t)$  (blue solid line) and $P_{ee\,0}(t)$ (red dashed line). Purity $\mu_q(t)$ [Panel(B)] and concurrence $C(t)$ [Panel (C)] relatively to the two-qubit subsystem.}\label{JointProb}
\end{center}
\end{figure}
\par
One of the main figure of merit to properly describe the unitary dynamics is the joint probability distribution for the energy states  $P_{Q\,n}(t)=\text{Tr}\left [ \rho(t) \ketbra{Q\,n}{Q\,n} \right]$. In Fig. \ref{JointProb}(A) we plot $P_{gg\,0}(t)$ and $P_{ee\,0}(t)$ showing the periodical behavior described above for which at times $t=t_1$ and $t=t_3$ the two-qubit subsystem exhibits maximally entangled Bell states [see, e.g., Eq. (\ref{2q_state_tn})] and the bosonic mode is in the vacuum state. We note that, depending on the initial state, only one of the two parity chains of states introduced before (see Fig. \ref{fig:ParityChains}) is involved in the dynamics, whereas the other one is not excited. For instance, in the case presented in Fig. \ref{JointProb}, the initial state is $\ket{gg\,0}$ and, thus, the dynamically excited states are those with parity $p=1$. Moreover, the plot of the purity  $\mu_q(t)=\text{Tr}[\rho^2_q(t)]$ of the two-qubit subsystem $\rho_q(t) = \text{Tr}_b[\rho(t)]$ [see panel (B) of Fig.~\ref{JointProb}] confirms the factorization of the states of the two subsystems at $t=t_n$ [for which $\mu_q(t_n)=1$] since the state of the whole system is always pure [see Eq.  (\ref{Psi_t_I})]. This dynamics, at $t=t_2$, also predicts a flip of both qubits,  which are found to be in the state $\ket{ee}$.
\par
Another useful tool for the analysis of the system dynamics is the concurrence,\cite{WootterConc} which measures the entanglement of two-qubit states and is defined as $C\equiv\text{max}\{0,\lambda_1-\lambda_2-\lambda_3-\lambda_4\}$ where $\lambda_i$ (with $i=1,\ldots, 4$) are the eigenvalues of the operator $R=\sqrt{\sqrt{\rho_q}\tilde{\rho_q}\sqrt{\rho_q}}$, with $\tilde{\rho_q}=(\sigma_y\otimes\sigma_y)\rho_q^*(\sigma_y\otimes\sigma_y)$ and $\rho_q^*$ the complex conjugate of the two-qubit density operator $\rho_q$. The concurrence is a bounded quantity $0\leq C \leq 1$ and in Fig. \ref{JointProb}(C) we show that only at the predicted times $t=t_1$ and $t=t_3$ the two qubits get maximally entangled.
\par
All these figures of merit will be used in the next section to fully describe the dynamics of the system beyond the slow-qubit approximation, in order to investigate the validity of the analytical model and the behavior of the strongly interacting system towards resonance conditions. 
\begin{figure}[t]
 \centering
\subfigure{\includegraphics[scale=0.5]{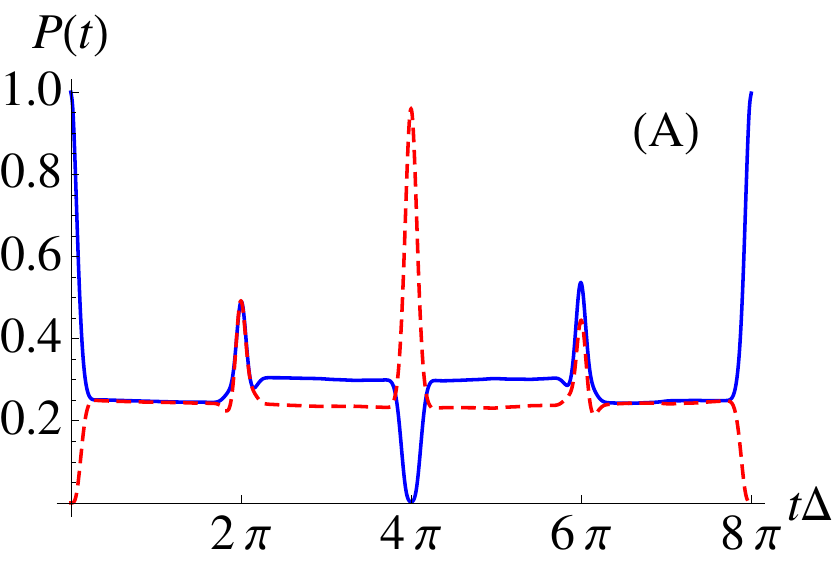}}
\hspace{5mm}
\subfigure{\includegraphics[scale=0.5]{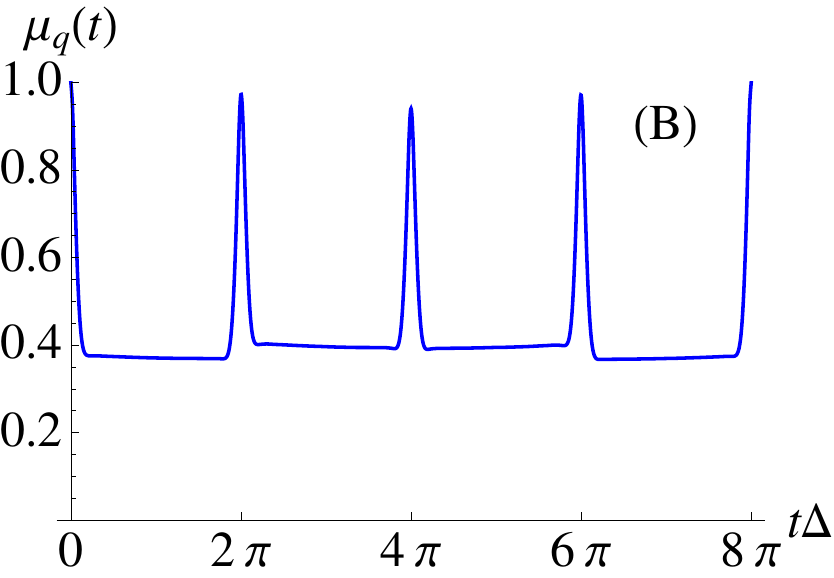}}\\
\subfigure{\includegraphics[scale=0.5]{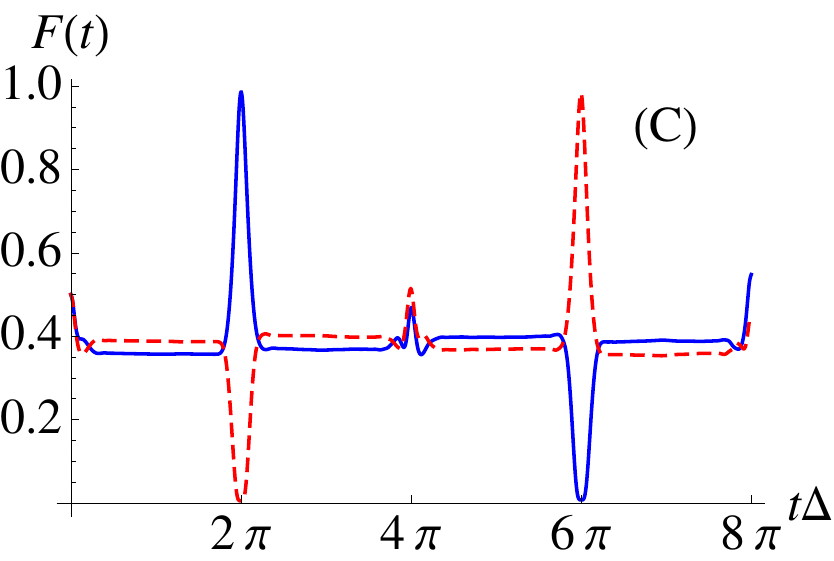}}
\hspace{5mm}
\subfigure{\includegraphics[scale=0.5]{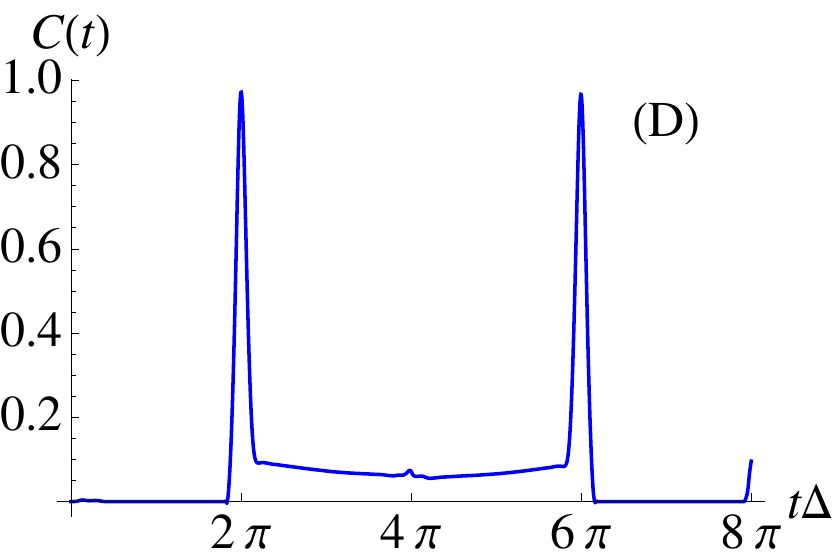}}
\caption{(Color online) Numerical integration for the system dynamics as a function of  $t\Delta$ starting from the state $\ket{gg\, 0}$, with parameters $\omega_q=0.1\omega$, $\Gamma_1=2\Delta$ and $\Gamma_2=\frac{65}{32}\Delta\approx2.03\,\Delta$. Panel (A): Joint probability distributions $P_{gg\,0}(t)$  (blue solid line) and $P_{ee\,0}(t)$ (red dashed line). Purity $\mu_q(t)$ [Panel(B)], fidelity with respect to the target states $\ket{\psi^{(+)}}$ (blue solid line) and $\ket{\psi^{(-)}}$ (red dashed line) and concurrence $C(t)$ [Panel (D)].}\label{fig:w0.1}
 \end{figure}

\section{Numerical analysis beyond the slow-qubit regime}\label{sec:numerical}
\begin{figure}[t]
\centering
\subfigure{\includegraphics[scale=0.5]{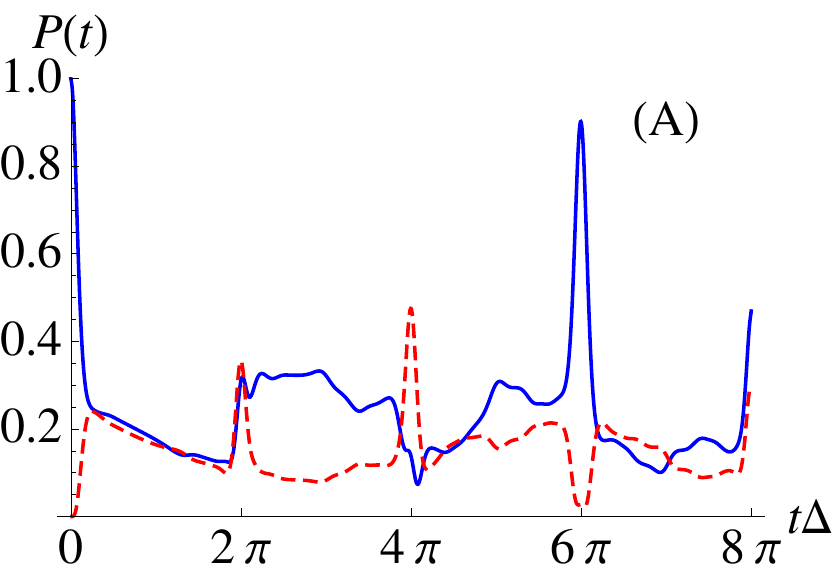}}
\hspace{5mm}
\subfigure{\includegraphics[scale=0.5]{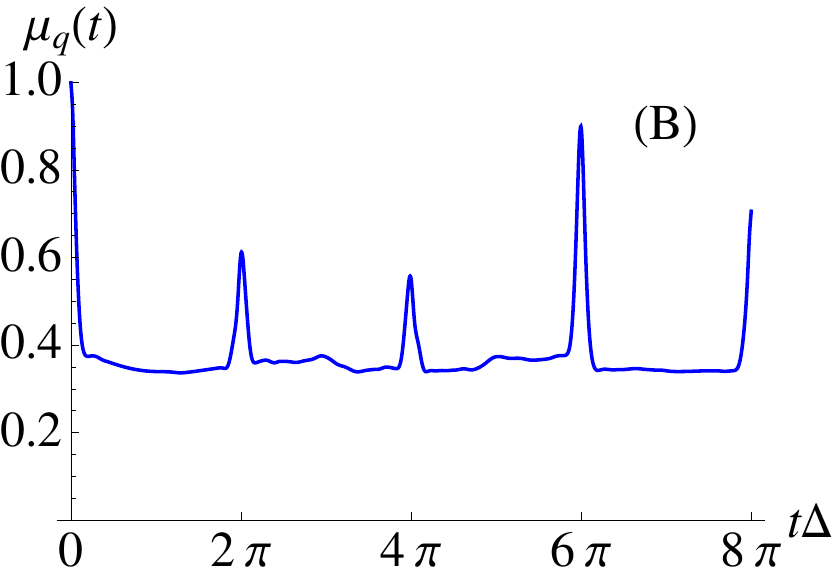}}\\
\subfigure{\includegraphics[scale=0.5]{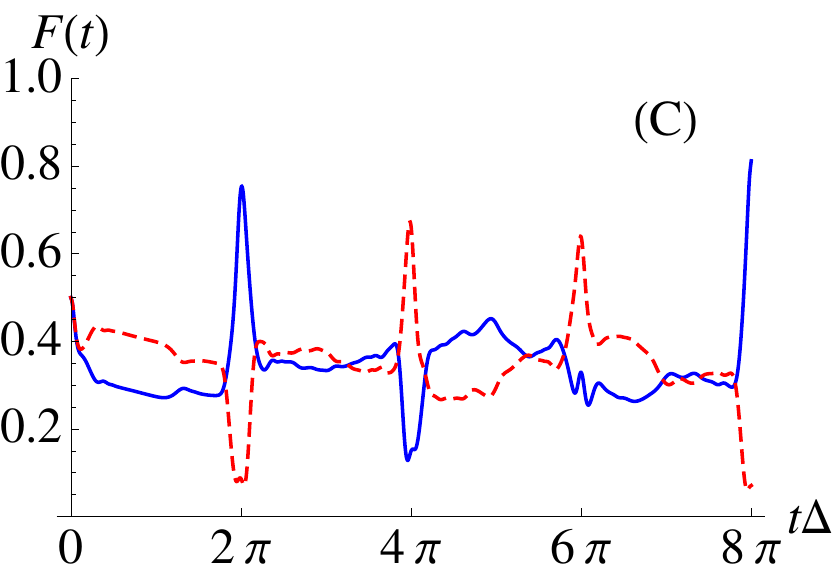}}
\hspace{5mm}
\subfigure{\includegraphics[scale=0.5]{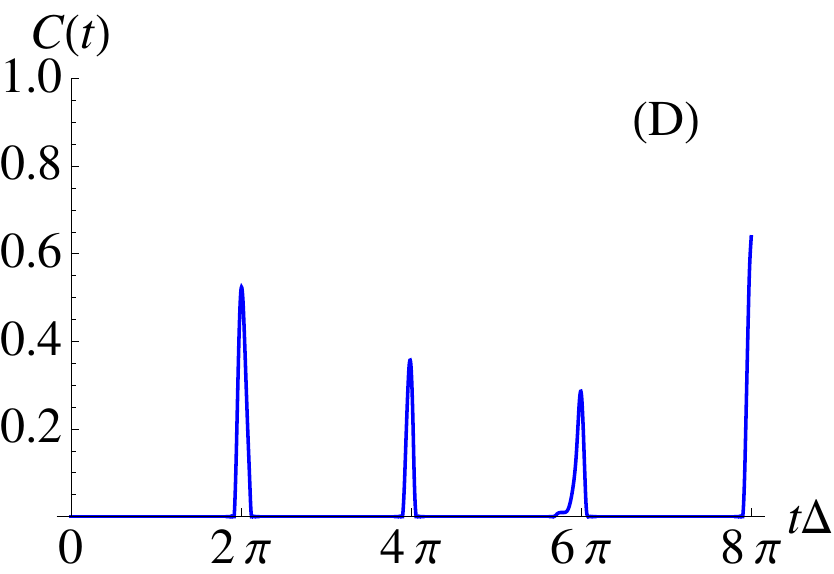}}
\caption{(Color online) Numerical integration for the system dynamics as a function of  $t\Delta$ starting from the state $\ket{gg\, 0}$, with parameters $\omega_q=0.5\,\omega$, $\Gamma_1=2 \Delta$ and $\Gamma_2=\frac{65}{32}\Delta \approx2.03\, \Delta$. Panel (A): Joint probability distributions $P_{gg\,0}(t)$  (blue solid line) and $P_{ee\,0}(t)$ (red dashed line). Purity $\mu_q(t)$ [Panel(B)], fidelity $F(t)$ [panel(C)] with respect to the target states $\ket{\psi^{(+)}}$ (blue solid line) and $\ket{\psi^{(-)}}$ (red dashed line) and concurrence $C(t)$ [Panel (D)].}\label{fig:w0.5}
 \end{figure}
Numerical integration of the system dynamics is needed as the Rabi model, in its full range of parameters, does not show an explicit analytic solution, though its integrability has been recently demonstrated\cite{prlBraak}. In this section we investigate a regime in which the qubit frequency is no longer negligible, exploring the behavior of the system close to the resonance condition $\omega_q\to\omega$, dropping the slow-qubit approximation and evolving the system from the Hamiltonian in Eq. (\ref{HRabi}). In particular we focus on the generation of maximally entangled qubit states aiming at finding a range of validity of the slow-qubit approximation, analyzing the figures of merit introduced in the previous section. In addition, we consider also the fidelity between the evolved state and one of the Bell states in Eq. (\ref{BellStates}), in order to verify the closeness, in the Hilbert state space, to the targeted entangled state. The numerical simulations of the system dynamics are performed for two choices of the qubits frequencies, in one case we consider equal transition frequencies, whereas in a second case we consider only one qubit in the slow-qubit regime.
\par
We begin our analysis from the first case. The increase of $\omega_q$ modifies the dynamics in such a way that the perfect oscillatory behavior described analytically before (see Fig. \ref{JointProb}) is distorted by the presence of the qubits, which acts as a dephasing term. Indeed, in Figs. \ref{fig:w0.1} and \ref{fig:w0.5} we show, for $\omega_q=10^{-1}\,\omega$ and $\omega_q=5\times 10^{-1}\,\omega$ respectively, how the oscillations for the generation of maximally entangled Bell states gets spoiled, in particular the joint probabilities [panel (A)] $P_{ee,0}(t)$ (blue solid line) and $P_{gg,0}(t)$ (red dashed line), the purity of the evolved two-qubit state [panel (B)], the fidelities $F(t)=\bra{\psi^{(\pm)}}\rho_q(t)\ket{\psi^{(\pm)}}$ [panel (C)] and the corresponding concurrence $C(t)$ [panel (D)].
\par
\begin{figure}[t]
 \centering
\includegraphics[scale=0.32]{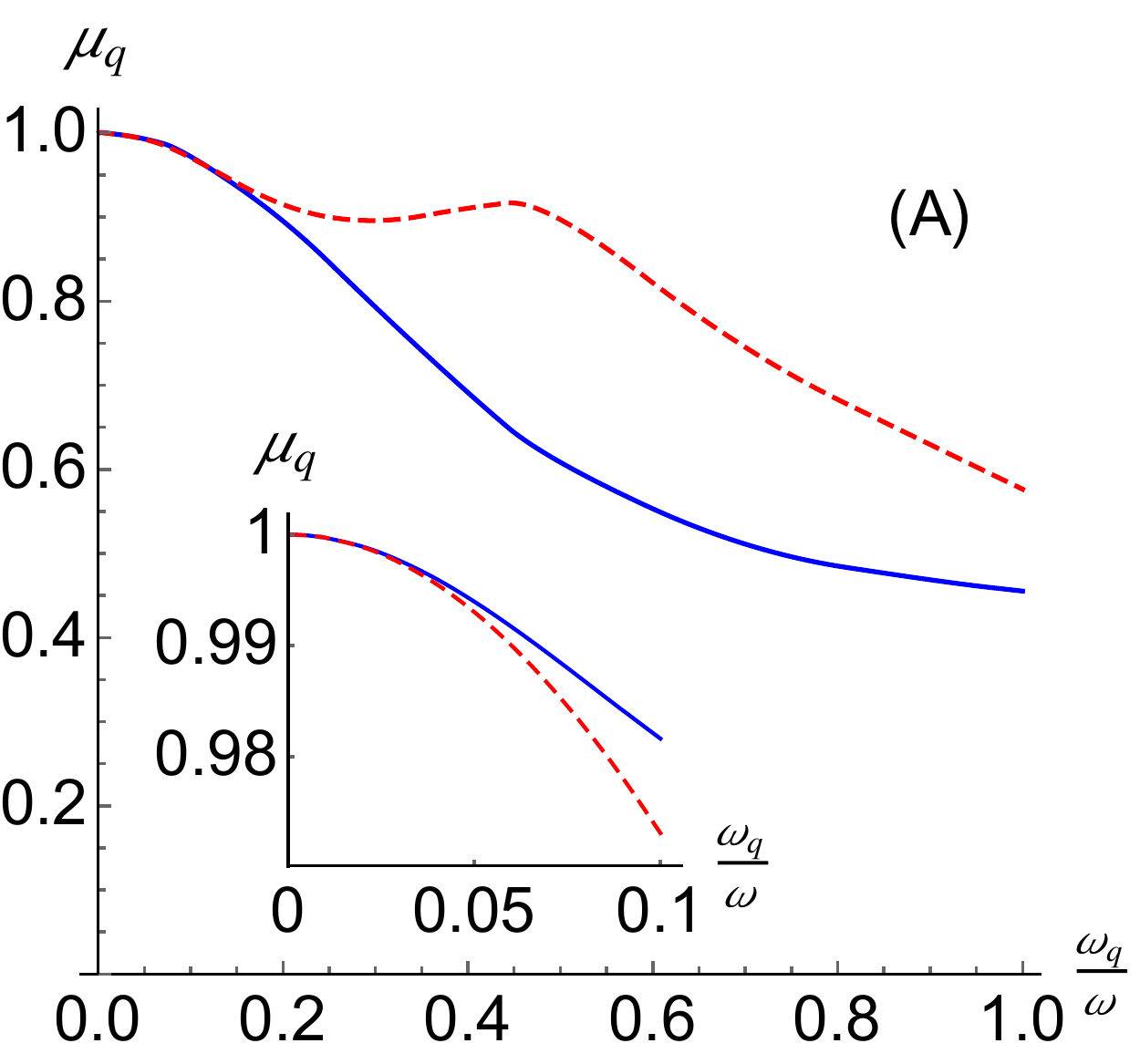}
\includegraphics[scale=0.32]{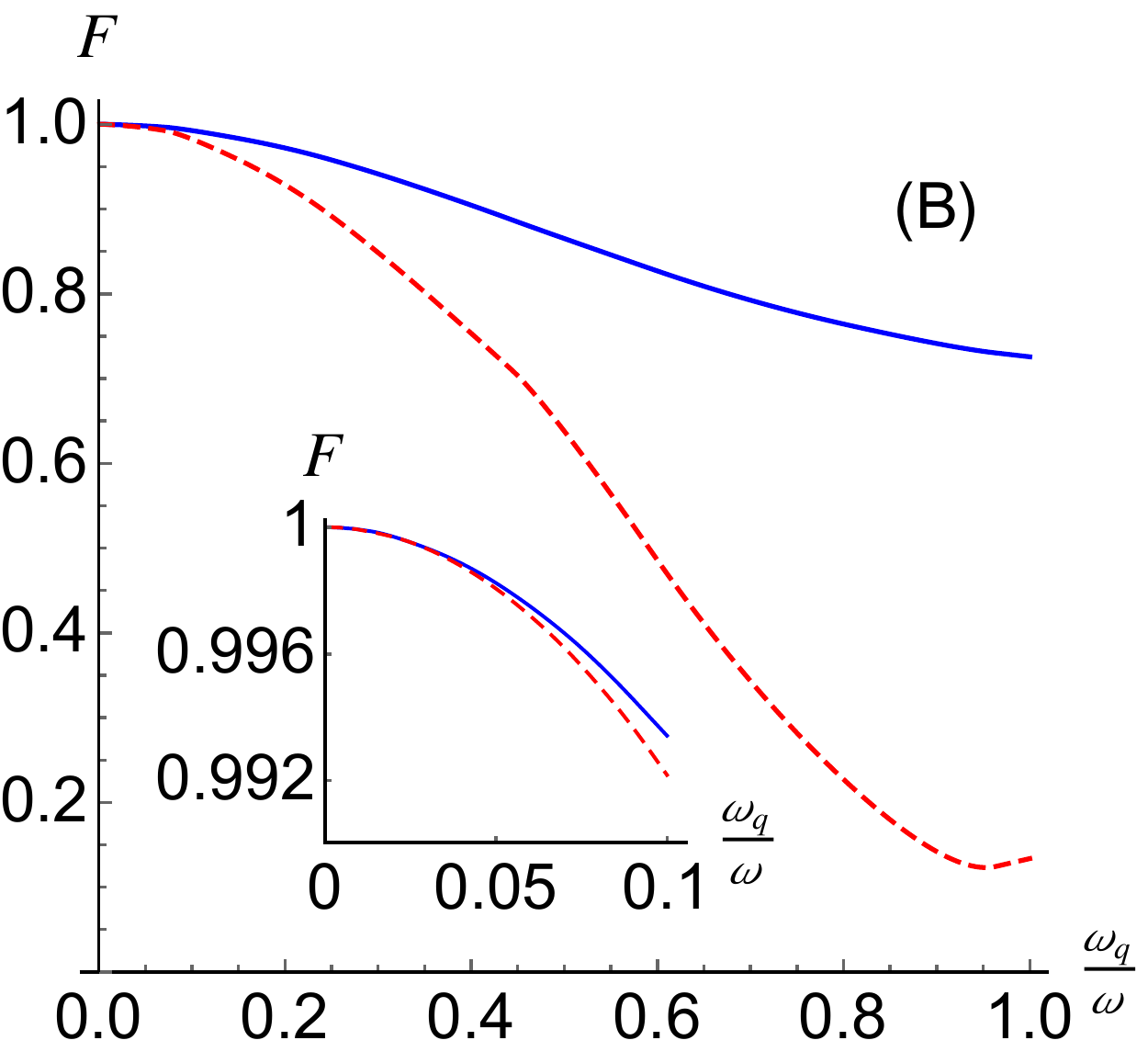}
\includegraphics[scale=0.32]{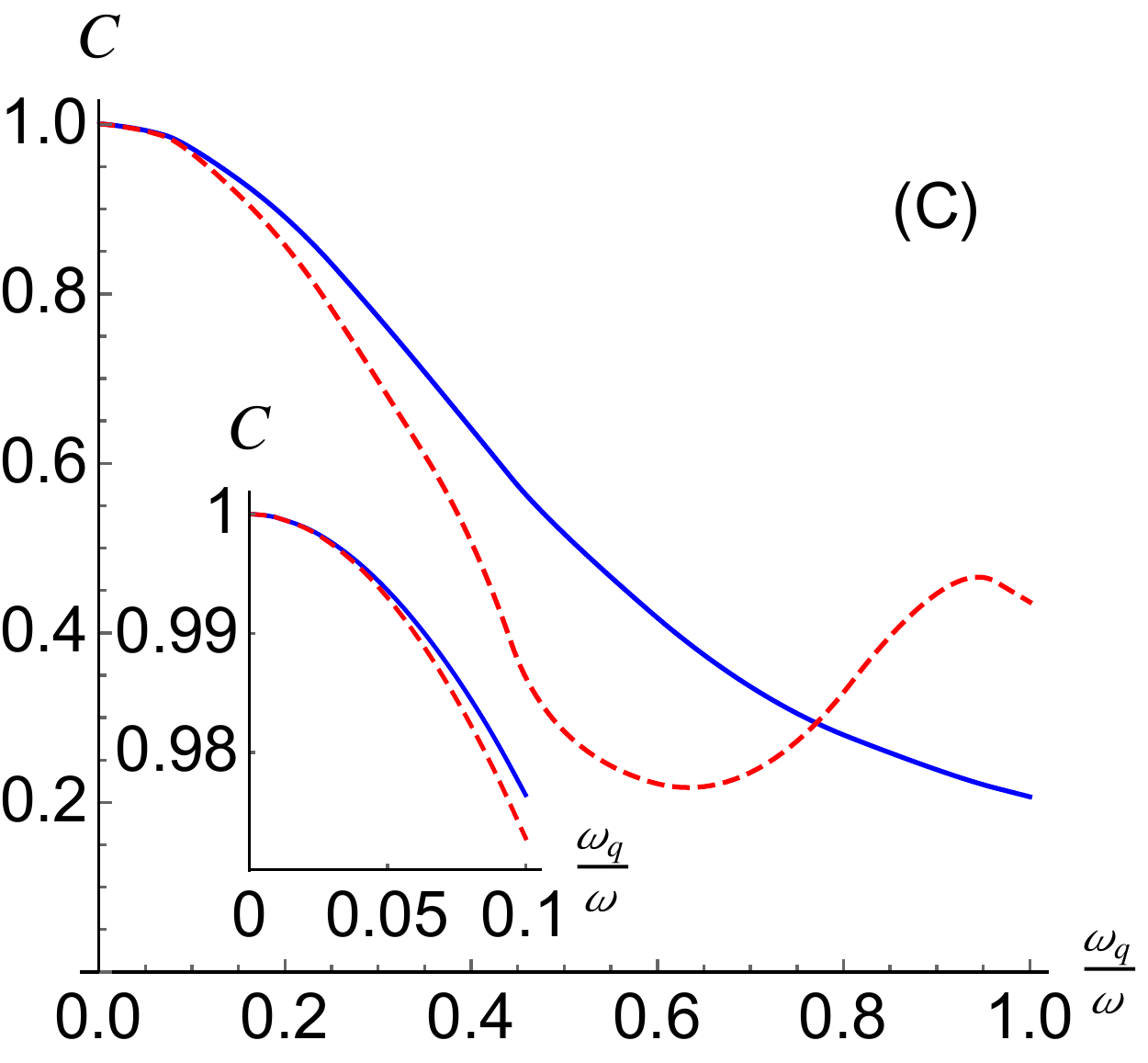}
 \caption{(Color online) Plot of purity (A), fidelity (B) and concurrence (C) at the peak times $t=t_1$ (blue solid line) and $t=t_3$ (red dashed line) corresponding to the target Bell states $\ket{\psi(t_1)}_{\rm{I}}$ and $\ket{\psi(t_3)}_{\rm{I}}$, as a function of the ratio $\omega_q/\omega=\omega_2=\omega_q$, with $\Gamma_1=2\Delta$ and $\Gamma_2=\frac{65}{32}\Delta\approx2.03\, \Delta$. The insets show the zoomed range $0<\omega_q<10^{-1}\,\omega$ where the analytical model and numerical results are in good agreement (high values of purity, fidelity and concurrence).}\label{fig:FreqWide}
 \end{figure}

In order to study the extension to this scenario of the analytical results of the previous sections we consider the behavior of the figures of merit, varying the qubit frequency, at the time peaks at which the Bell states $\ket{\psi(t_1)}_{\rm{I}}$ and $\ket{\psi(t_3)}_{\rm{I}}$ are generated. In Fig. \ref{fig:FreqWide} we report the purity, the fidelity and the concurrence of the two-qubit subsystem as a function of the qubit frequency $\omega_q$ for values covering the entire range up to the resonance condition $\omega_q=\omega$. The insets of these plots show the range of validity of the slow-qubit approximation, say $0<\omega_q\leq 10^{-1}\,\omega$, according to the high values of fidelity to the target Bell states, confirmed also by high values of the concurrence which measures the actual entanglement between the two qubits.\cite{bina:fid}
\par
\begin{figure}[t]
\subfigure{\includegraphics[scale=0.32]{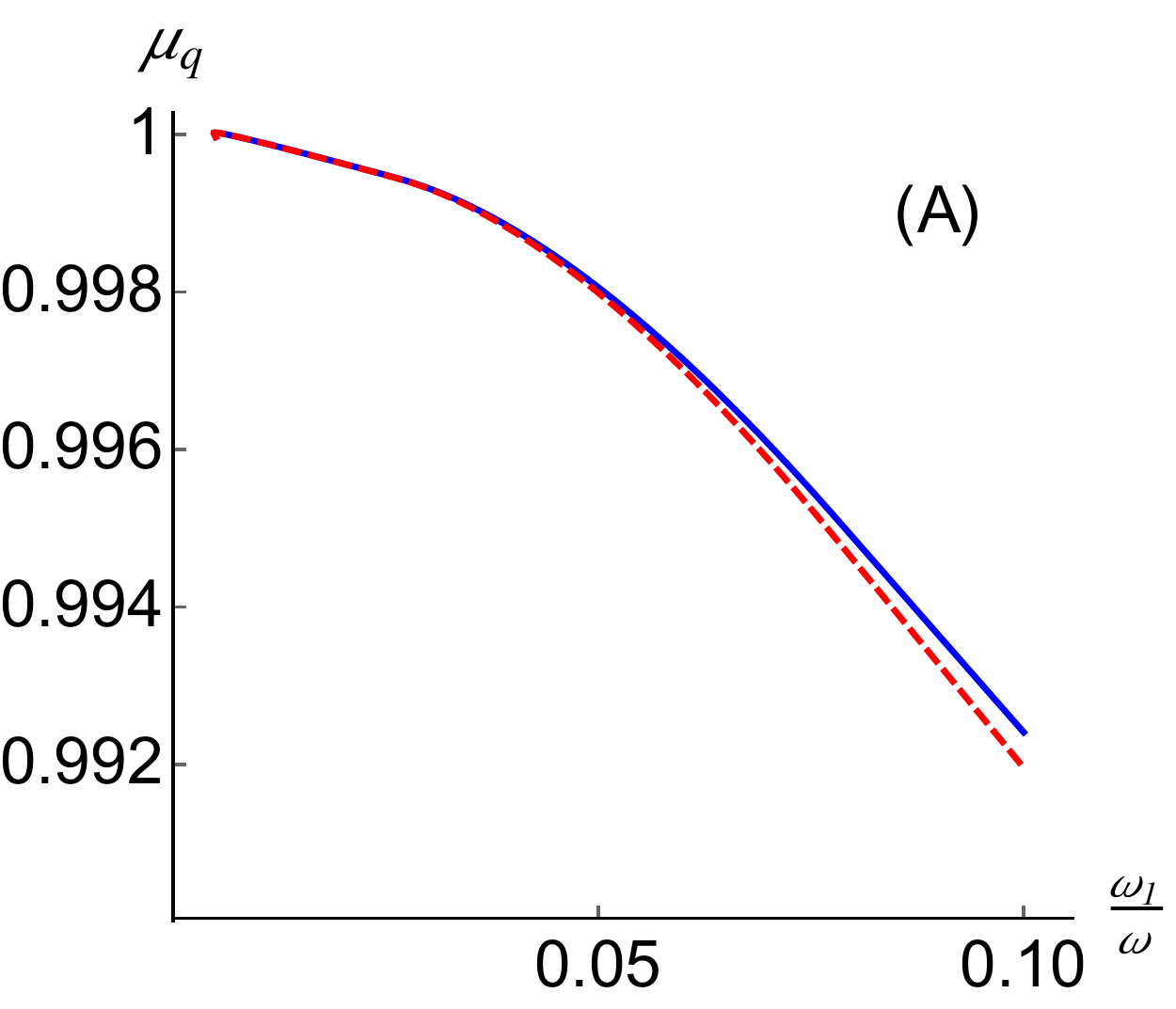}}
\subfigure{\includegraphics[scale=0.32]{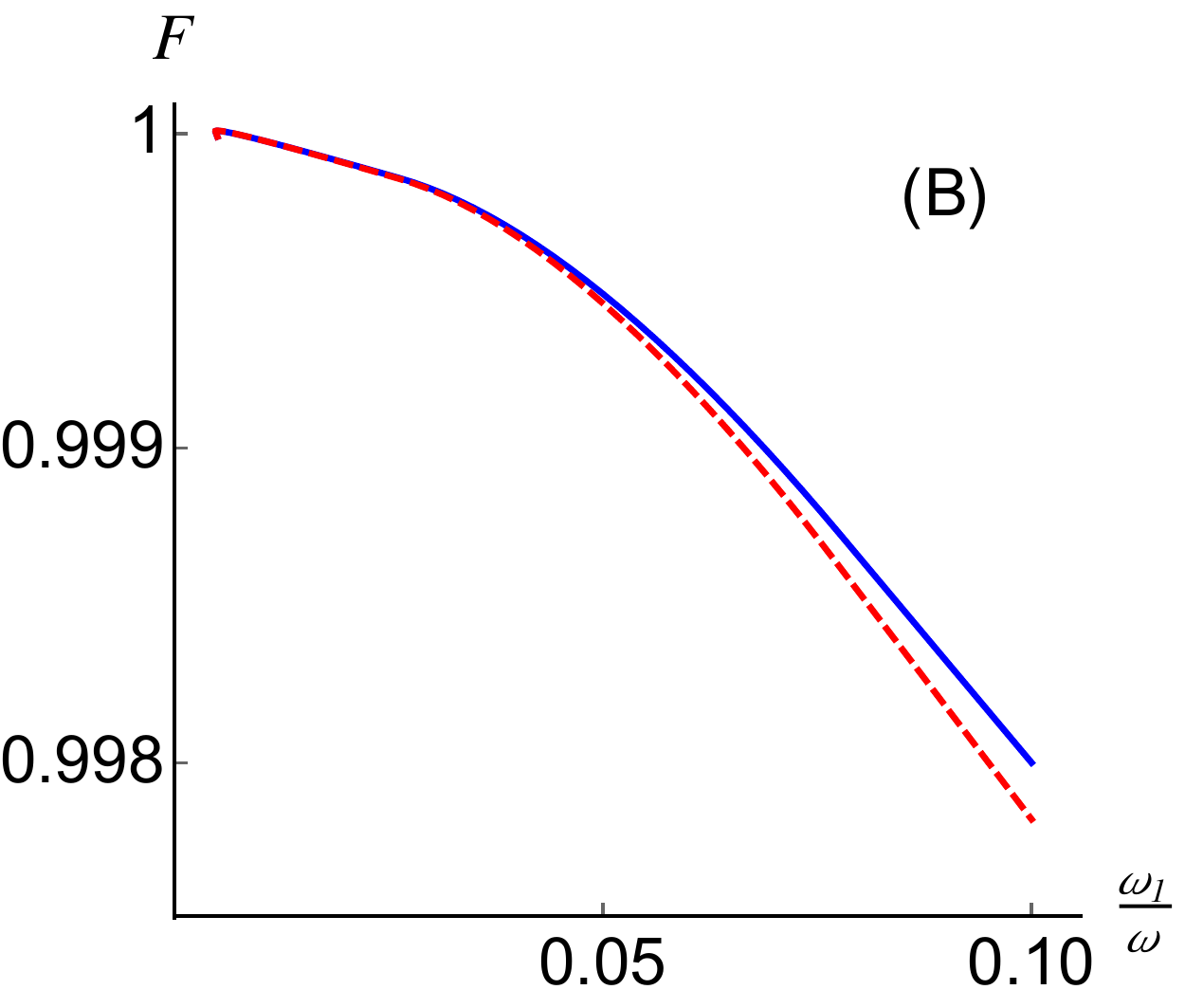}}
\subfigure{\includegraphics[scale=0.32]{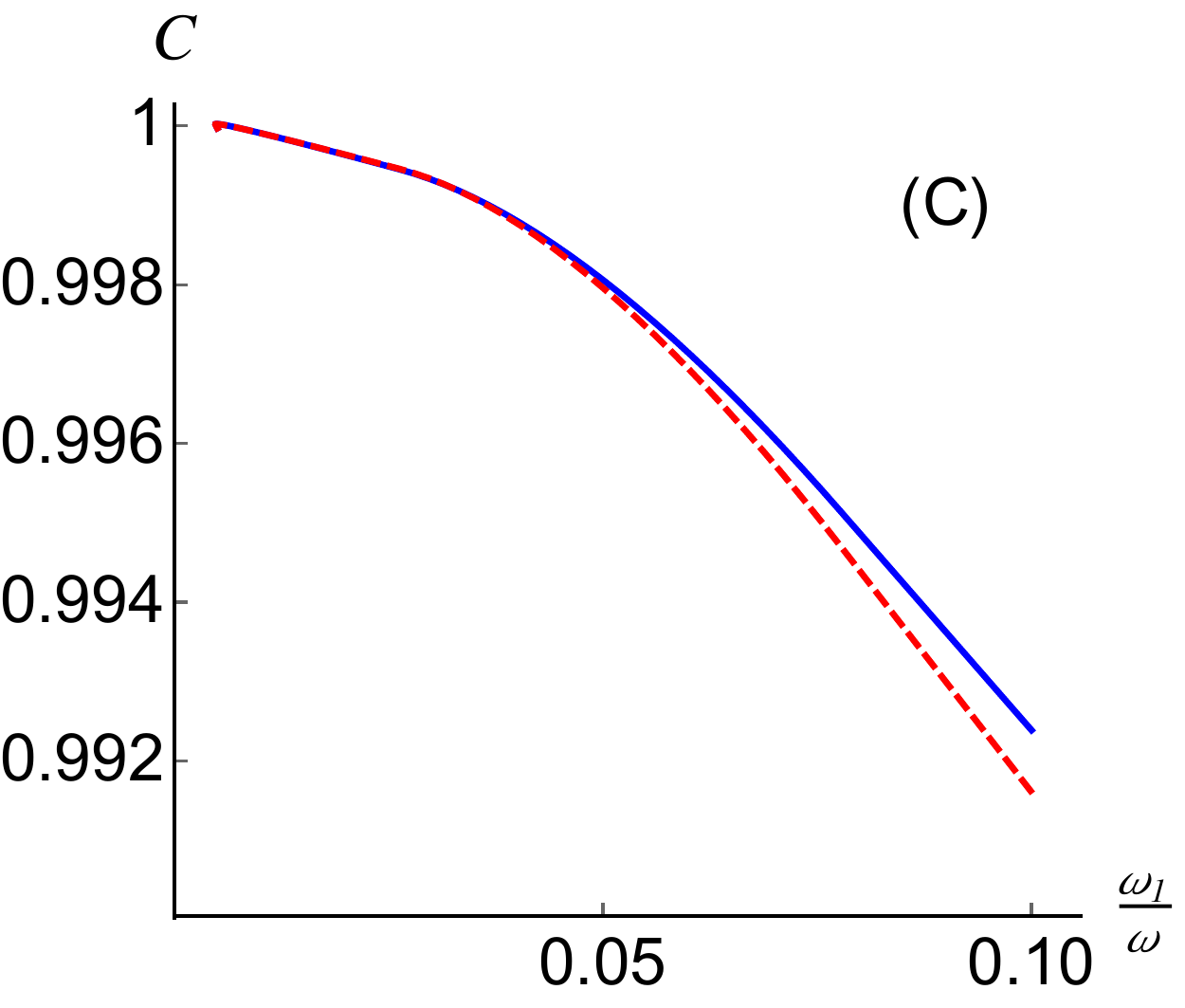}}
\caption{(Color online) Plot of purity (A), fidelity (B) and concurrence (C) at the peak times $t=t_1$ (blue solid line) and $t=t_3$ (red dashed line) corresponding to the target Bell states $\ket{\psi(t_1)}_{\rm{I}}$ and $\ket{\psi(t_3)}_{\rm{I}}$, as a function of the ratio $\omega_q/\omega$, with $\Gamma_1=2\Delta$ and $\Gamma_2=\frac{65}{32}\Delta\approx2.03\, \Delta$. The transition frequency of the second qubit is fixed at $\omega_2=10^{-3}\,\omega$.}\label{fig:AFreqNarrow}
 \end{figure}

A further analysis concerns the aforementioned case in which two qubits with different transition frequencies are considered, especially when one of the two fulfills the slow-qubit approximation. We performed numerical simulations fixing $\omega_2=10^{-3}\,\omega$ and varying $\omega_1$ towards the resonance condition $\omega_1=\omega$. We show in Fig. \ref{fig:AFreqNarrow} the behavior of  purity, fidelity and concurrence for the two-qubit subsystem in the range $0<\omega_q<10^{-1}\,\omega$, which is the same considered in the previous case (equal qubit frequencies). Remarkably, this choice of the parameters enhances the performances in the generation of Bell states at the first time peaks, as fidelity and concurrence display higher values than the previous case.

\section{Conclusions}\label{sec:concl}
We investigated the quantum Rabi model for two qubits with values of the coupling constants to a single mode resonator belonging to the deep strong coupling regime. Considering low transition frequencies for the two qubits, we performed the slow-qubit (or degenerate) approximation studying the problem of a quantized oscillator conditionally displaced by the state of the qubits. The physical dynamics of the system captured by the slow-qubit approximation in the deep strong coupling regime is radically different from a JC-like dynamics, as it involves multi-photon processes and in principle spans the entire Fock space of the resonator mode. The oscillatory behavior of the dynamics generates maximally entangled Bell states, depending on a precise relation between the coupling constants. We explored the range of validity of this approximation by means of numerical integration of the Schr\"odinger equation ruled by the whole quantum Rabi Hamiltonian, to study at which extent the qubit frequencies can be increased to obtain a high-fidelity generation of Bell states. Even though only the ultra-strong coupling regime up to $\Gamma\simeq 3\times 10^{-1}\,\omega$ has been reached recently, we believe that circuit QED is the most promising setting for the breakthrough of the deep strong coupling regime and the realization of ultrafast quantum gates. Nonetheless, quantum simulations for effective Hamiltonians mimicking the quantum Rabi model for two qubits or other experimentally inaccessible models are currently the best strategy to attack these kind of problems.

\section*{Acknowledgments}
M.B. is grateful to E. Solano, G. Romero and D. Braak for fruitful discussions. M.B. and S.O. acknowledge A. Lulli for the stimulating collaboration. This work has been supported by MIUR (FIRB ``LiCHIS'' Ñ RBFR10YQ3H).

\end{document}